# On the origin of the premature breakdown of thermal oxide on 3C-SiC probed by electrical scanning probe microscopy

P. Fiorenza[1], E. Schilirò[1], F. Giannazzo[1], C. Bongiorno[1], M. Zielinski[2], F. La Via[1], F. Roccaforte[1]

[1] Consiglio Nazionale delle Ricerche – Istituto per la Microelettronica e Microsistemi (CNR-IMM), Strada VIII, n.5 Zona Industriale, 95121 Catania, Italy

[2] NOVASiC, Savoie Technolac, BP267, F-73375 Le Bourget-du-Lac Cedex, France

patrick.fiorenza@imm.cnr.it

**Abstract**. The dielectric breakdown (BD) of thermal oxide ($SiO_2$) grown on cubic silicon carbide (3C-SiC) was investigated comparing the electrical behavior of macroscopic metal-oxide-semiconductor (MOS) capacitors with nanoscale current and capacitance mapping using conductive atomic force (C-AFM) and scanning capacitance microscopy (SCM). Spatially resolved statistics of the oxide BD events by C-AFM revealed that the extrinsic premature BD is correlated to the presence of peculiar extended defects, the anti-phase boundaries (APBs), in the 3C-SiC layer. SCM analyses showed a larger carrier density at the stacking faults (SFs) the 3C-SiC, that can be explained by a locally enhanced density of states in the conduction band. On the other hand, a local increase of minority carriers concentration was deduced for APBs, indicating that they behave as conducting defects having also the possibility to trap positive charges. The results were explained with the local electric field enhancement in correspondence of positively charged defects.

## Introduction

Thanks to its unique physical and electronic properties, silicon carbide (SiC) is an excellent material for high power and high temperature electronics [1]. Among the different polytypes of SiC, the most mature is the hexagonal 4H-SiC, which is currently available on large area (commercially up to 150 mm, 200mm for R&D substrates) and suitable for the fabrication of power electronics devices. On the other hand, the cubic polytype (3C-SiC) has been studied since a long time, as it could give some advantages with respect to 4H-SiC in metal oxide semiconductor field effect (MOSFET) devices. In fact, due to its smaller band gap, the $SiO_2$/3C-SiC interface states density at energies close to the conduction band edge is expected to be much lower than that typically measured for $SiO_2$/4H-SiC interfaces [2,3,4]. Consequently, a high inversion channel mobility is expected in 3C-SiC MOS-based devices. As a matter of fact, in the last decade 3C-SiC MOSFETs with excellent mobility values (50-260 $cm^2V^{-1}s^{-1}$) have been demonstrated in literature [5,6,7,8]. Hence, 3C-SiC is considered a good candidate for power electronics applications in the voltage range 600-900V.

Another potential advantage of the cubic polytype is the possibility to be hetero-epitaxially grown on large diameter cheap silicon (Si) substrates. However, the large lattice mismatch results in 3C-SiC layers with a high density of defects and a notable surface roughness [9]. Hence, both the oxidation process and the final quality of the $SiO_2$/3C-SiC interfaces will be strongly dependent on the 3C-SiC material [10,11,12,13].

In literature, thermally grown and deposited oxides have been investigated on 3C-SiC [14,15,16]. In both cases, the $SiO_2$/3C-SiC system often suffers of a premature dielectric breakdown, and is characterized by a high interface state density and large negative shift of the flat band voltage ($V_{FB}$) [17], indicating the presence of a net positive effective charge in the system. Recently, *Li et al.* [18] observed that the dielectric breakdown kinetics of MOS capacitors is influenced by the gate area and they proposed the existence of at least three different extrinsic and intrinsic breakdown mechanisms responsible for the MOS degradation. However, the origin of each extrinsic mechanism and its relation with the crystalline defects are still unclear.

Clearly, due to the aforementioned issues, MOS-based devices on 3C-SiC materials are still far from a practical application in power electronics.

In this work, the mechanisms that induce a premature breakdown of thermally grown $SiO_2$ on hetero-epitaxial 3C-SiC/Si layers were investigated by means of both measurement on MOS capacitors and nanoscale electrical characterizations. The origin of the different premature dielectric breakdown behavior was pursued at the nanoscale by conductive atomic force microscopy (C-AFM) and scanning capacitance microscopy (SCM). Particular attention was put on the impact of the 3C-

SiC crystalline defects, like stacking faults (SFs) and anti-phase boundaries (APBs), as they are also known to affect the electrical behavior of Schottky contacts on 3C-SiC [19,20].

**Experimental section**

For this investigation a 10.2 µm thick 3C-SiC layer has been grown on on-axis Si(100) substrates by chemical vapour deposition (CVD) using silane ($SiH_4$) and propane ($C_3H_8$) as silicon and carbon precursors, respectively [21].

The morphology of the oxidized 3C-SiC surface was investigated using a PSIA XE-150 AFM operating in non-contact mode with highly doped silicon tips.

On the oxidised material, lateral capacitors were fabricated with the following procedure. After a sacrificial oxidation and oxide removal to clean deeply the 3C-SiC surface, a thermal oxide was grown at 1150°C for one hour in dry $O_2$. Lateral metal-oxide-semiconductor (MOS) capacitors were obtained by defining Ni/Au metal electrodes by means of photolithography and lift-off. In particular, the electrodes consisted in a circular inner gate electrode, 50 µm of radius, surrounded by a large area metal large electrode. With this geometry, the capacitance of the cathode is sufficiently high with respect of that of the inner electrode and can be neglected, as they are connected in series. The capacitance-voltage (C-V) and the current-voltage (I-V) characteristics of the MOS capacitors were measured in a CASCADE Microtech probe station, using a Keysight B1505A parameter analyzer.

Nanoscale electrical characterizations, namely Conductive Atomic Force Microscopy (C-AFM) and Scanning Capacitance Microscopy (SCM), have been performed using a DI3100 system by Bruker with a Nanoscope V controller, equipped with the TUNA module in the case of the current measurements using Bruker boron doped diamond tips that provide a tip contact diameter < 100 nm.

Transmission electron microscopy (TEM) images were collected at 200 kV by a JEOL 2010F microscope equipped with the Gatan imaging filter.

**Results and discussion**

The material characterization started with the evaluation of surface morphology and the identification of the electrical active defects in the as-grown 3C-SiC before thermal oxidation. In particular, C-AFM was employed to probe the 3C-SiC bare surface conduction as schematically illustrated in Figure 1a. In such configuration, the C-AFM probe slides on the 3C-SiC surface acting as a nano-Schottky contact biased in forward polarization with respect to the silicon substrate working as a large back-contact. Figure 1b and 1c show the surface morphology and the current map, respectively, collected simultaneously with a tip bias of $V_{tip} = + 0.5V$. Under this polarization, the current map (Figure 1c) shows the presence (in an appropriate current range from 0 to 20 pA) of preferential conductive paths with a current level at least one order of magnitude higher than the surrounding 3C-SiC bare material. This observation is consistent with the one reported in a recent C-AFM study of 3C-SiC on silicon, where the nature of the extended conductive defects was elucidated by cross-comparison with atomic resolution structural analyses and ab-initio simulations [20]. In particular, the boundaries between two Anti-Phase 3C-SiC domains were identified by dotted curved lines in Figures 1b and c, and are commonly named Anti-Phase Boundaries (APBs). These are extended defects separating 3C-SiC regions with inverted crystal symmetry, that is, upside-down flipping of the Si-C bond. The APBs appear as curved and randomly oriented features in the C-AFM map. On the other hand, the conductive features appearing as straight lines in Figure 1c were identified as some stacking faults (SFs). These are typical planar crystallographic defects lying on {111} planes of the 3C-SiC polytype, which are terminated as lines with mutually perpendicular orientations on the (100) growth plane.

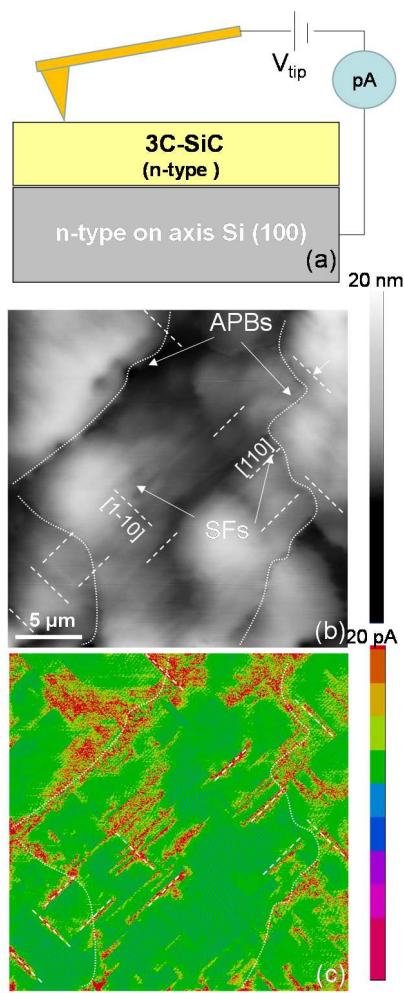

Figure 1: a) Schematic illustration of the setup for the electrical characterization of the 3C-SiC surface by C-AFM. b) Morphology and c) current map collected under forward bias (+0.5 V) of the tip. Both SFs and APBs are visible and more conductive than the surrounding 3C-SiC material.

Figure 2a shows an AFM image of the 3C-SiC material after the thermal growth of SiO$_2$. The morphology of the as-grown oxide resembles that of the 3C-SiC material, i.e., characterized by *terraces* separated by darker lines associated with the APBs. The SiO$_2$ surface roughness, evaluated in terms of the root mean square of the heights distribution (RMS), was in the order of 5 nm. This value is comparable to those measured directly on 3C-SiC layer before oxidation, demonstrating the conformal growth of the oxide onto the substrate. Figures 2b and c show two representative cross sectional TEM micrographs collected on the regions where a SF (b) and an APB (b) reach the 3C-SiC/SiO$_2$ interface. A conformal oxide growth, without any thickness variations at defects positions, can be clearly observed.

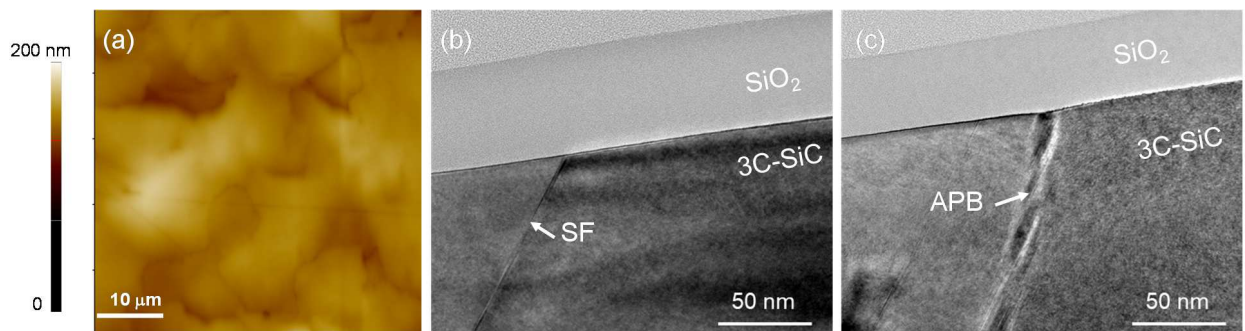

Figure 2. AFM morphology image of the thermal SiO$_2$ grown on 3C-SiC (a) and cross-sectional TEM micrographs in the SF region (b) and in the APB region (c) respectively. No thickness variation of the SiO$_2$ layer can be observed.

On this material, lateral MOS capacitors were fabricated according to the schematic shown in Figure 3a. Figure 3b shows the optical microscope image of the top of the MOS capacitors with the indication of the gate (anode) area separated by the cathode.

Firstly, the electrical quality of the oxide was evaluated by means of capacitance - voltage (C-V) and current density – voltage (J-V) measurements on MOS capacitors. Figure 3c shows the experimental C-V curve compared with the theoretical one. From the value of the depletion capacitance, the net doping concentration of the 3C-SiC layer was estimated to be $N_D = 2\times10^{16}$ cm$^{-3}$.

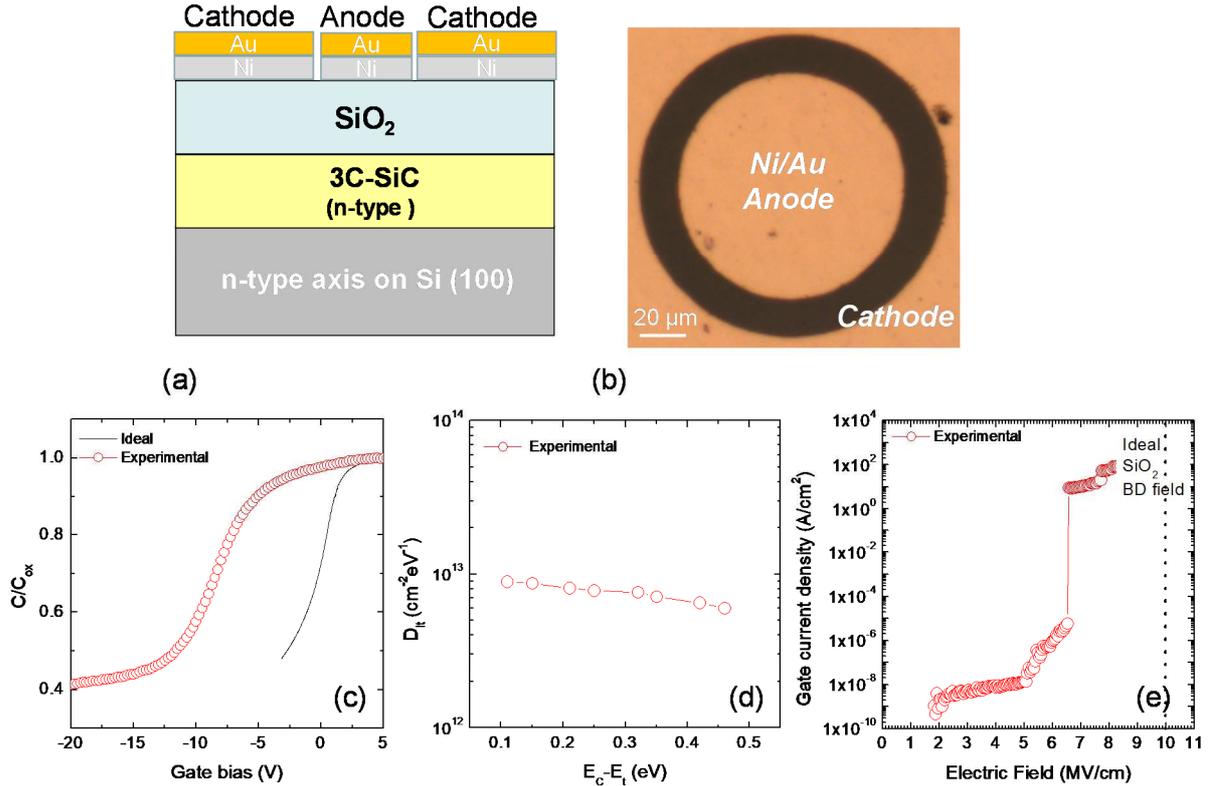

Figure 3: (a) Schematic in cross-section of the lateral MOS capacitor fabricated on thermally oxidized 3C-SiC; (b) optical microscope image of a MOS capacitor; (c) Experimental C-V curve measured on 3C-SiC MOS capacitors compared with the theoretical curve; (d) Experimental interface state density energetic distribution; (e) Current density vs voltage (J-V) curve measured on 3C-SiC MOS capacitors. The value of the ideal $SiO_2$ breakdown electric field is indicated by the dashed line.

The oxide thickness, estimated from the accumulation capacitance ($C_{ox}$), is about 40 nm. The experimental C-V curve is negatively shifted with respect to the ideal C-V curve and has a flat band voltage ($V_{FB}$) of about – 7.5 V (Figure 3c). The observed $V_{FB}$ negative shift (Fig. 3a) corresponds to an amount of positive effective charge in the MOS system of $N_{eff}= +3\times10^{12}$ cm$^{-2}$. A similar negative flat band voltage shift has been observed by *Sharma et al.* [15] on thermal oxides on 3C-SiC grown in the 1200-1400°C range. In particular, they showed that the $V_{FB}$ shift increased with increasing the oxidation temperature [15]. The origin of the positive effective charge is still debated, and has been attributed to the presence of carbon clusters, positively or negatively charged O-H-C-Si complexes, and dangling bonds formed after thermal oxidation [22].

The interface states density ($D_{it}$) energy profiles below the 3C-SiC conduction band edge, estimated using the conductance method, are depicted in Fig. 3d. The maximum value, measured at 0.1 eV below the conduction band, was $8\times10^{12}$ cm$^{-2}$eV$^{-1}$. These $D_{it}$ values are in the same order of magnitude of the typical values reported for thermal oxides on 3C-SiC [15,16].

Finally, Fig. 3e shows the J-V curve collected on the lateral MOS capacitor. As can be seen, the breakdown field is significantly lower with respect to the theoretical value for $SiO_2$ (~10 MV/cm) [23]. In fact, the experimental BD field is about 6.5 MV/cm.

To get a deeper insight on the origin of this discrepancy, a nanoscale electrical characterization of the $SiO_2$/3C-SiC system will be presented in the remaining part of this paper.

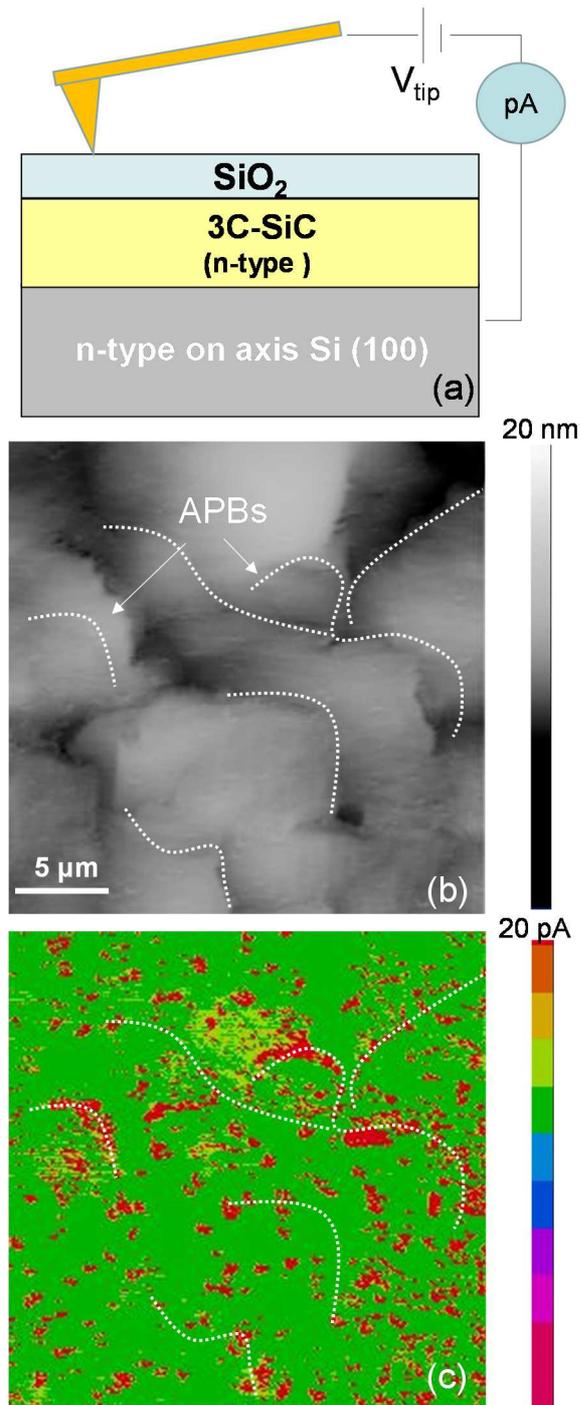

Figure 4: a) Schematic illustration of the setup for the electrical characterization of the $SiO_2$/3C-SiC system by C-AFM. b) Morphology and c) current map collected under high electric field (8 MV/cm) of the tip.

Nanoscale resolution current mapping by C-AFM [24,25,26] is a powerful method to investigate the role played by defects on the dielectric breakdown behavior of thin insulators [27]. In order to explain the origin of the premature breakdown observed at the macroscopic level in MOS capacitors, thin thermal oxides (~ 10 nm) were characterized at a nanoscale level by means of C-AFM stress measurements at very high electric field (> 8 MV/cm), i.e., close to the ideal breakdown field value of silicon dioxide (~10 MV/cm) and above the experimental breakdown field (~6.5 MV/cm) measured on macroscopic $SiO_2$/3C-SiC capacitors (Figure 3e). Figure 4a shows a schematic illustration of the experimental setup used for nanoscale C-AFM analyses of the $SiO_2$/3C-SiC system. Figures 4b and 4c report simultaneously collected AFM morphology and current map by the positively bias conductive diamond tip with $V_{tip}$=8 V. The C-AFM current map in Fig. 4c can be considered as a breakdown map of an array of nano-MOS capacitors (with areas corresponding to the tip contact area) simultaneously stressed at 8 MV/cm for each nano-device. Then, the position of the BD spots in red in Figure 4c have been correlated with the surface morphology (Figure 4b). As can be seen, the BD events are not randomly distributed but there are some regions with a larger density compared with the surroundings. Drawing a dashed line to guide the eye it is possible to see the correspondence on the morphology (Figure 4b) of weak lines located in correspondence with some APBs, already identified in the current maps of the bare 3C-SiC surface (Figure 1c). Noteworthy, the straight lines conductive features associated to SF in the bare 3C-SiC surface were not visible in the case of 3C-SiC covered by thermal $SiO_2$. It can be also noticed that the edges of the 3C-SiC crystalline terraces seem to have no influence on the nano-MOS premature breakdown.

C-AFM maps allowed to identify the weak points responsible of premature breakdown. In addition, accelerated stress tests of the thin $SiO_2$ layer could be performed at the nanoscale, by varying the stress time applied by the tip according to the experimental procedure illustrated in Refs. [28,29]. More specifically, the stress time for each nano-MOS was varied by changing the scan rate of the tip (i.e., the scan time per line T that can be varied from 1 s up to 10 min) and the image scan size (i.e., the scan line length L that can be varied from 200 nm up to 100 µm). In this way the stress time applied by the biased conductive diamond tip, expressed as $t_{stress}$= *T/(L/a)* with *a*≈100 nm the tip diameter, was varied over about 3 orders of magnitude. The amount of nano-MOS that survived to a given stress time was evaluated from the current maps acquired at the different stress times. Hence, such a variable time stress measurements enabled to count the cumulative failure rate according to the Weibull statistics [30].

As an example, a comparison between two sequential current maps collected in the same position (namely the same array of nano-MOS) is shown in Figure 5. In particular, Figures 5a and 5b show the current maps collected by the C-AFM stressing each nano-MOS at fixed bias for 2s and 50s, respectively. Figure 5c shows the comparison between the distribution of the current flowing through all the nano-MOS after stress time of 2s and 50s, respectively. As can be noticed, for the 2s stress time, the current values distribution (at 8 MV/cm) is mainly centered at the bottom of the sensitivity of the C-AFM (100 fA) and only few nano-MOS capacitors show current values in the 12 pA range. On the other hand, for the 50s stress time, the current values distribution (at 8 MV/cm) is bimodal with a peak at 100 fA and a sharp peak at current larger than 12 pA (the upper sensitivity value of the C-AFM). Hence, in the same map there are nano-MOS where the current value is suddenly increased up to three orders of magnitude, which can be identified as the breakdown events.

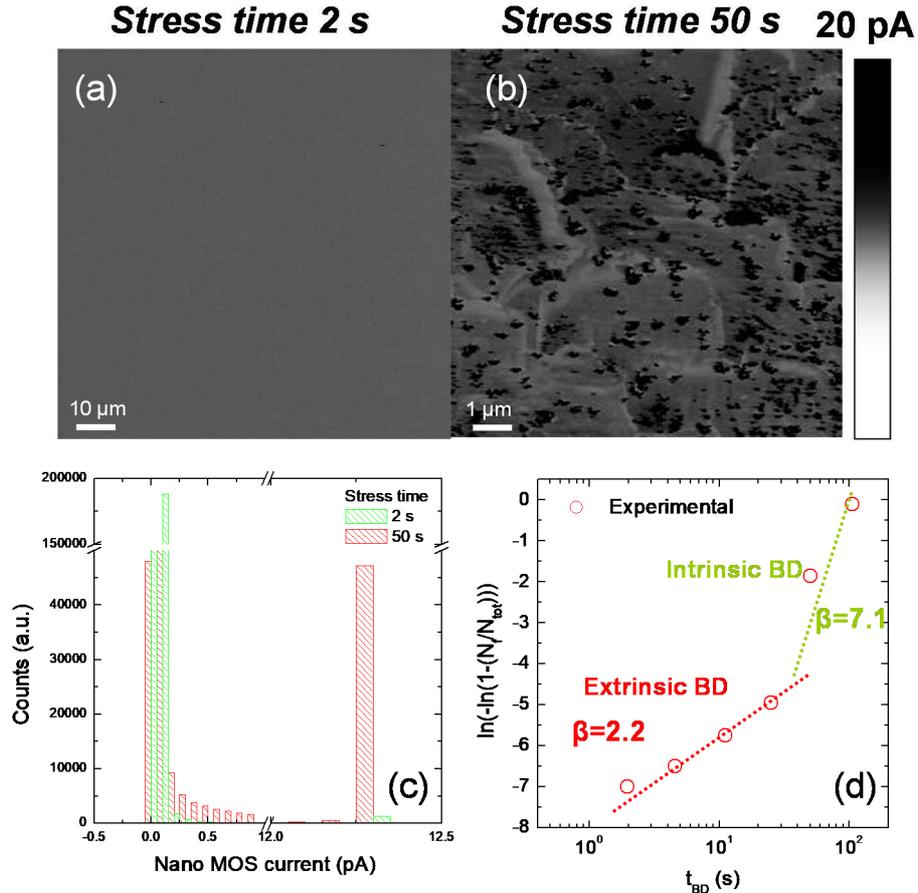

Figure 5: C-AFM current maps acquired under high electric field (8 MV/cm) of the tip with a stress time of 2s (a) and 50s (b); (c) Distributions of the current values flowing in the nano-MOS after stress times of 2s and 50s; (d) Breakdown kinetics and Weibull plots. The intrinsic and extrinsic BD mechanisms are highlighted by dashed lines.

Fig. 5d shows the Weibull's plot obtained after stressing the total amount ($N_{tot}$) of nano-MOS array for different times. As described in Fig. 4, the number (N) of failed nano-MOS increases increasing the stress time. At low values of the stress time, the Weibull plot has a low slope β = 2.2. This value is smaller than expected for a 10 nm thick insulator layer [31]. This result demonstrates the presence of an extrinsic breakdown mechanism that induces the premature failure of the nano-MOS. However, as already seen in other works [12,24], at high stress time it is possible to identify the regions of the insulator having a nearly ideal breakdown mechanism with a Weibull slope β = 7.1 (Fig. 5d). This behavior is in agreement with the prediction of the percolation theory [32] for a 10 nm thick $SiO_2$ layer.

The combination between the experimental results presented in Figures 1, 4 and 5 suggest that both SFs and APBs are conductive extended defects in the 3C-SiC material, but only APBs are the responsible of the premature BD of the SiO$_2$/3C-SiC nano-MOS structures. In fact, the breakdown events could be not correlated to the geometry of the SFs (Figure 1c).

Using C-AFM local stress measurements, *Kozono et al.* [33] correlated the BD events at SiO$_2$/4H-SiC interfaces with the step-bunching edges on the semiconductor surface, attributing the premature BD to the local electric field crowding on these morphological features.

Our results shown in Figure 2 and in Figure 4 suggest that the 3C-SiC morphological features have only a marginal impact on the BD kinetics of the nano-MOS. On the other hand, the impact of the APBs is prominent on the BD kinetics. Hence, the origin of the premature BD in correspondence of the APBs has to be pursued in the electronic nature of the defect and its impact on the MOS system.

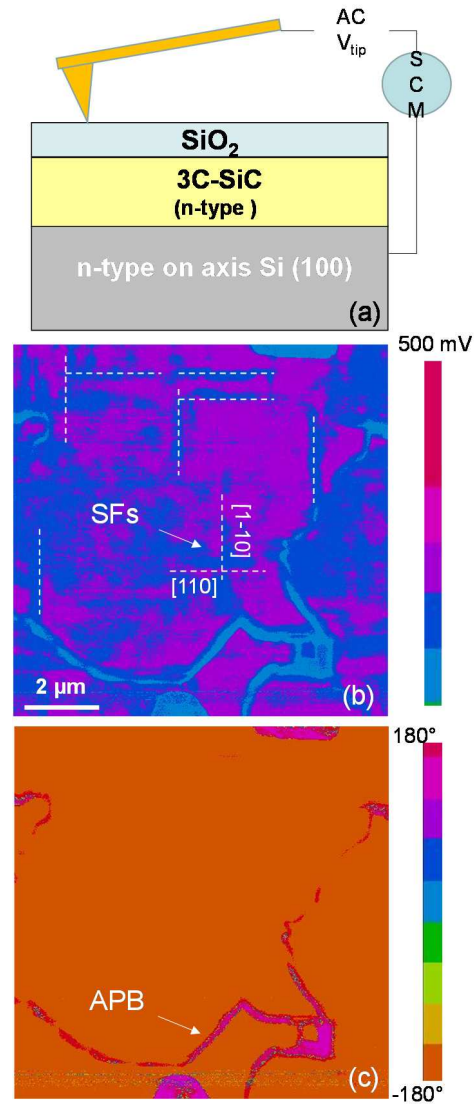

Figure 6: (a) Schematic illustration of the setup for the electrical characterization of the SiO$_2$/3C-SiC system by SCM; (b) SCM amplitude and (c) SCM phase collected under low signal (< 2 V$_{AC}$) of the tip.

In order to further elucidate the role of the extended defects in 3C-SiC on the electrical behaviour of the SiO$_2$/3C-SiC MOS capacitor, capacitance measurements were performed also at nanoscale [34] using the SCM technique. A schematic representation of the SCM experimental setup is illustrated in Figure 6a. During the SiO$_2$ surface scan with the diamond tip, an AC modulating bias at 100 kHz frequency and with amplitude $\Delta V=2V$ (below the conduction regime through the insulator) was applied to the sample, and the capacitance variation $\Delta C$ in response to this modulation was recorded with the SCM sensor. Figure 6b and 6c show two representative images of the SCM signal amplitude $|\Delta C|$ and of the phase signal. In particular, both straight line features resembling SFs (dashed lines) and curved line features resembling APBs can be distinguished in the SCM amplitude image in Figure 6b, whereas only APBs features can be observed in the phase map in Figure 6c. Lateral variations of the SCM signal amplitude in a nanoMOS capacitors can be related to local changes in the insulator thickness/permittivity or in the semiconductor depletion region underneath the tip (related to the local carrier density) [35]. However, since a uniform and conformal SiO$_2$ layer was shown by cross-sectional TEM analyses for the SiO$_2$/3C-SiC system under investigation (see Figures 2b and 2c), it can be concluded that the spatial variations in the SCM amplitude are related to local changes in the carrier density in the 3C-SiC layer. In particular, the lower $|\Delta C|$ values measured in Figure 6b on SFs as compared to the surrounding material suggests a locally higher carrier density.

Although the donor concentration incorporated in the 3C-SiC layer during the growth is expected to be uniform, the presence of defects (such as SFs and APBs) is known to be responsible of modifications of the 3C-SiC electronic band-structure with respect to defects' free regions [20], ultimately resulting in a change of the local carrier density in the 3C-SiC material. In particular, in the presence of a SF a significant enhancement of the density of states (DOS) both in the valence and the conduction band of 3C-SiC has been predicted by ab-initio calculations [20]. This can explain both the enhanced SFs conductivity observed in the C-AFM maps on the bare 3C-SiC surface (see Figure 1c) and the larger electron density deduced from the SCM amplitude map (see Figure 6b). Hence, SFs can be considered as highly conducting 2D defects.

Interestingly, Figures 6c shows a strong change of the SCM phase signal from 180° on APBs to -180° on the surrounding regions. The phase signal is known to be very sensitive to the type of majority carriers in the semiconductor. In the present case, the variation of the SCM phase can be also indicative of a local increase of the minority carriers (holes) concentration in the 3C-SiC material. Similar results have been recently obtained by SCM phase analyses performed on threading dislocations of 4H-SiC [36], which served to demonstrate an increase of minority carriers concentration in the volume surrounding these killer defects responsible of BD of 4H-SiC power MOSFETs after prolonged stress [36]. In the specific case of 3C-SiC, ab-initio calculations indicated that a peculiar property of APBs is that these defects introduce states within the 3C-SiC bandgap close to the valence band [20]. Thus, the APBs may act also as a preferential sites for positive charges trapping. The presence of positively charged defects at the SiO$_2$/3C-SiC interface is also in agreement with the negative flat band shift observed in the macroscopic MOS structures (Figure 2c).

In this scenario, an impact of the APBs in the accelerated BD kinetics can be argued. In fact, the accelerated BD occurred when a positive electric field is applied to the nano-MOS. Figure 7a shows the band gap diagram under the application of an electric field to the ideal nano-MOS. For a given potential applied to the gate ($V_{tip}$) the actual electric field value depends on the insulator thickness ($t_{ox}$), on the flat-band voltage ($V_{FB}$) and on the amount of effective positive charge ($N_{eff}$) in the nano-MOS:

$$E_{ox} = \frac{V_{tip} - V_{FB}}{t_{ox}} + q\frac{N_{eff}}{t_{ox}}$$

where q is the elementary electron charge.

Clearly, for a given $V_{tip}$ on the gate of the nano-MOS, the electric field is minimum in the ideal case (Figure 7a), while it is increased in the presence of positive charge trapping in the APBs (Figure 7b). The local increase of the electric field in correspondence of the APBs produces the increase of the local current flow related to the increase of the injected electrons from the 3C-SiC substrate into the insulating layer. Hence, the increased local electron injection accelerate the dielectric breakdown, producing extrinsic breakdown events. A similar role of charged interfacial defects has been invoked by *Arvanitopoulos et al*. [37] to describe the anomalous experimental current behaviour through Schottky barriers on 3C-SiC. Accordingly, the charged defects at the metal/3C-SiC interface induce an electrostatic thinning of the Schottky barrier, resulting in an enhanced current conduction.

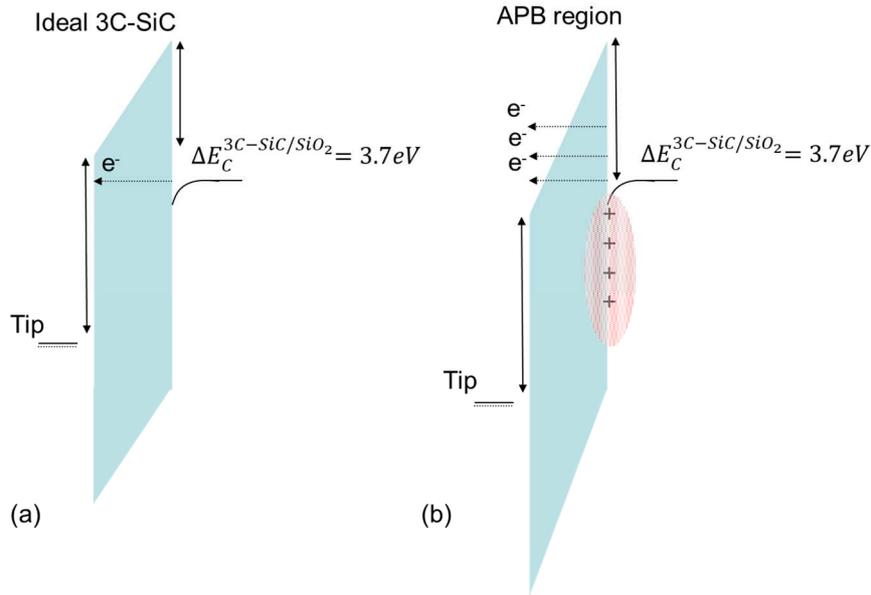

Figure 7: Schematic band diagram of the SiO$_2$/3C-SiC nano-MOS, when a high tip bias is applied (a) in the ideal material and (b) in the APB region. The positive charge trapped in the APB produces an increase of the electric field that enhances the tunnelling current from electrons injected from the substrate into the insulator accelerating its degradation.

**Conclusion**

In this paper, the origin of the premature breakdown behavior of thermal oxide grown on 3C-SiC/Si hetero epitaxial layers has been investigated, employing electrical scanning probe microscopy techniques and standard characterization on large area MOS capacitors. In particular, the nanoscale Weibull statistic plot obtained by C-AFM on the nano-MOS revealed a non-negligible population of extrinsic breakdown events. Those extrinsic BD events are correlated to the presence of anti-phase boundaries (APBs) in the 3C-SiC. SCM characterization on the oxide allowed to demonstrate that the stacking faults (SFs) possess a larger density of states in the 3C-SiC conduction band. On the other hand, the SCM phase signal in proximity of the APBs was interpreted with the presence of positive trapped charge. Basing on these nanoscale results, the premature dielectric breakdown observed in large area MOS capacitors has been explained interpreted by the presence of positively charged APBs, which cause an enhanced electron injection from the semiconductor into the insulator. This model agrees with the localization of the extrinsic BD in correspondence of the APBs.

**Acknowledgements**

This work has been supported by the European project CHALLENGE (Call: H2020-NMBP-2016-2017, grant. agreement 720827).

The authors would like to thank S. Monnoye and H. Mank (Novasic) for the 3C-SiC growth and polishing, and S. Di Franco (CNR-IMM) for the MOS capacitor fabrication.

**References**

[1] T. Kimoto, J. Cooper, Fundamentals of Silicon Carbide Technology: Growth, Characterization, Devices and Applications, John Wiley & Sons, Singapore Pte. Ltd. (2014).

[2] K.K. Lee, G. Pensl, M. Soueidan, G. Ferro, Y. Monteil, Very Low Interface State Density From Thermally Oxidized Single-Domain 3C–SiC/6H–SiC Grown by Vapour–Liquid–Solid Mechanism, Jpn. J. Appl. Phys. 45, (2006) 6823

[3] A. Schöner, M. Krieger, G. Pensl, M. Abe, H. Nagasawa, Fabrication and Characterization of 3C-SiC-Based MOSFETs, Chem. Vapor Depos. 12, (2006) 523

[4] P. Fiorenza, F. Giannazzo, F. Roccaforte, Characterization of SiO2/4H-SiC Interfaces in 4H-SiC MOSFETs: A Review, Energies 12, (2019) 2310

[5] H. Nagasawa, M. Abe, K. Yagi, T. Kawahara, N. Hatta, Fabrication of high performance 3C-SiC vertical MOSFETs by reducing planar defects, phys. stat. sol. (b) 245, (2008) 1272

[6] M. Kobayashi, H. Uchida, A. Minami, T. Sakata, R. Esteve, A. Schöner, 3C-SiC MOSFET with High Channel Mobility and CVD Gate Oxide, Mater. Sci. Forum 679-680, (2011) 645

[7] F. Li, Y. Sharma; D. Walker; S. Hindmarsh; M. Jennings; D. Martin, C. Fisher, P. Gammon, A. Pérez-Tomás, P. Mawby, 3C-SiC Transistor with Ohmic Contacts Defined at Room Temperature, IEEE Electron Device Lett. 37, (2016) 1189

[8] K. K. Lee, Y. Ishida; T. Ohshima; K. Kojima; Y. Tanaka; T. Takahashi; H. Okumura et al, K. K. Lee, Y. Ishida; T. Ohshima; K. Kojima; Y. Tanaka; T. Takahashi; H. Okumura, K. Arai, T. Kamija, IEEE Electron Dev. Lett., 24, (2003) 466–468

[9] F. La Via, A. Severino, R. Anzalone, C. Bongiorno, G. Litrico, M. Mauceri, M. Schoeler, P. Schuh, P. Wellmann, From thin film to bulk 3C-SiC growth: Understanding the mechanism of defects reduction, Mater. Sci. Semicond. Proc. 78, (2018) 57

[10] M. Eickhoff, N. Vouroutzis, A. Nielsen, G. Krötz, J. Stoemenos, Oxidation Dependence on Defect Density in 3C-SiC Films, J. Electroch. Soc. 148, (2001) G336.

[11] A. Severino, M. Camarda, S. Scalese, P. Fiorenza, S. Di Franco, C. Bongiorno,A. La Magna, and F. La Via, Preferential oxidation of stacking faults in epitaxial off-axis (111) 3C-SiC films, Appl. Phys. Lett. 95, (2009) 11905

[12] J. Eriksson, F. Roccaforte, P. Fiorenza, M.-H. Weng, F. Giannazzo, J. Lorenzzi, N. Jegenyes, G. Ferro, V. Raineri, Nanoscale probing of dielectric breakdown at SiO2 /3C-SiC interfaces, J. Appl. Phys. 109, (2011) 013707

[13] R. Anzalone, S. Privitera, A. Alberti, N. Piluso, P. Fiorenza and F. La Via, Electrical properties evaluation on high quality hetero-epitaxial 3C-SiC(001) for MOSFET applications, Mater. Sci. Forum, 821-823, (2015) 773

[14] R. Esteve, A. Schöner, S. A. Reshanov, C.-M. Zetterling, H. Nagasawa, Comparative study of thermally grown oxides on n-type free standing 3C-SiC (001), J. Appl. Phys. 106, (2009) 044513

[15] Y.K. Sharma, F. Li, M. R. Jennings, C. A. Fisher, A. Pérez-Tomás, S. Thomas, D. P. Hamilton, S. A. O. Russell, P. A. Mawby, High-Temperature (1200–1400°C) Dry Oxidation of 3C-SiC on Silicon, J. Electron. Mater. 44, (2015) 4167

[16] R. Anzalone, S. Privitera, M. Camarda, A. Alberti, G. Mannino, P. Fiorenza, S. Di Franco, F. La Via, Interface state density evaluation of high quality hetero-epitaxial 3C–SiC(001) for high-power MOSFET applications, Mater. Sci. Eng. B 198, (2015) 14

[17] K. Cherkaoui, A. Blake. Y. Y. Gomeniuk, J. Lin, B. Sheehan, M. White, P. K. Hurley, P. J. Ward; Investigating positive oxide charge in the SiO2/3C-SiC MOS system, AIP Adv. 8, (2018) 085323

[18] F. Li, Q. Song, A. Perez-Tomas, V. Shah, Y. Sharma, D. Hamilton, C. Fisher, P. Gammon , M. Jennings, P.Mawby A First Evaluation of Thick Oxide 3C-SiC MOS Capacitors Reliability, IEEE Trans. Electr. Dev., 67, (2020) 237

[19] J. Eriksson, F. Roccaforte, F. Giannazzo, R. Lo Nigro, V. Raineri, J. Lorenzzi, G. Ferro; Improved Ni/3C-SiC contacts by effective contact area and conductivity increases at the nanoscale, App. Phys. Lett. 94 (2009) 112104

[20] F. Giannazzo, G. Greco, S. Di Franco, P. Fiorenza, I. Deretzis, A. La Magna, C. Bongiorno, M. Zimbone, F. La Via, M. Zielinski, F. Roccaforte ; Impact of Stacking Faults and Domain Boundaries on the Electronic Transport in Cubic Silicon Carbide Probed by Conductive Atomic Force Microscopy Adv. Electron. Mater., 6 (2020) 1901171

[21] M. Zielinski, S. Monnoye, H. Mank, C. Moisson, T. Chassagne, A. Michon, M. Portail., Structural Quality, Polishing and Thermal Stability of 3C-SiC/Si Templates, Mater. Sci. Forum 924 (2018) 306

[22] X. Shen and S.T. Pantelides, Identification of a major cause of endemically poor mobilities in SiC/SiO2 structures, Appl. Phys. Lett. 98 (2011) 053507

[23] P. Fiorenza, R. Lo Nigro, V. Raineri, D. Salinas ; Breakdown kinetics at nanometer scale of innovative MOS devices by conductive atomic force microscopy, Microelectr. Eng. 84 **2007** 441

[24] Conductive Atomic Force Microscopy: Applications in Nanomaterials (Ed: M. Lanza), Wiley-VCH, Weinheim, Germany (2017).

[25] Electrical Atomic Force Microscopy for Nanoelectronics (Ed: U. Celano), Springer Nature, Switzerland (2019).

[26] F. Hui, M. Lanza, Scanning probe microscopy for advanced nanoelectronics, Nat. Electron.*,* 2 (2019) 221

[27] M Porti, M Nafrıa, X Aymerich, A Olbrich, B Ebersberger ; Electrical characterization of stressed and broken down SiO2 films at a nanometer scale using a conductive atomic force microscope, J. Appl. Phys. 91 (2002) 2071

[28] P. Fiorenza, R. Lo Nigro, V. Raineri, S. Lombardo, R. G. Toro, G. Malandrino, I. L. Fragalà, Breakdown kinetics of Pr2O3 films by conductive-atomic force microscopy, App. Phys. Lett. 87 (2005) 231913

[29] P. Fiorenza, V. Raineri, Reliability of thermally oxidized SiO2/4H-SiC by conductive atomic force microscopy, Appl. Phys. Lett. 88 (2006) 212112

[30] D.K. Schroder, Semiconductor Material and Device Characterization, 3rd ed.;Wiley: Hoboken, NJ, USA, (2006).

[31] P. Fiorenza, R. Lo Nigro, V. Raineri, D. Salinas, Conductive Atomic Force Microscopy Studies on the Reliability of Thermally Oxidized SiO2/4H-SiC, Mater. Sci. Forum. 556-557 (2007) 501

[32] R. Degraeve, G. Groeseneken, R. Bellens, J. L. Ogier, M. Depas, P. J. Roussel, H.E. Maes: New insights in the relation between electron trap generation and the statistical properties of oxide breakdown; IEEE Trans. Electron. Dev. 45 (1998) 904

[33] K. Kozono, T. Hosoi, Y. Kagei, T. Kirino, S. Mitani, Y. Nakano, T. Nakamura, T. Shimura, H. Watanabe, Direct Observation of Dielectric Breakdown Spot in Thermal Oxides on 4H-SiC(0001) Using Conductive Atomic Force Microscopy, Mater. Sci. Forum 645–648 (2010) 821

[34] P. Fiorenza, R. Lo Nigro, V. Raineri, R.G. Toro M.R. Catalano; Nanoscale imaging of permittivity in giant-k CaCu3Ti4O12 grains, J. Appl. Phys. 102 (2007) 116103

[35] F. Giannazzo, D. Goghero, and V. Raineri, Experimental aspects and modeling for quantitative measurements in scanning capacitance microscopy, J. Vac. Sci. Technol. *B* 22 (2004) 2391-2397

[36] P. Fiorenza, M. S. Alessandrino, B. Carbone, C. Di Martino, A. Russo, M. Saggio, C. Venuto, E. Zanetti, F. Giannazzo, F. Roccaforte; Understanding the role of threading dislocations on 4H-SiC MOSFET breakdown under high temperature reverse bias stress, Nanotechnology 31 (2020) 125203

[37] A. E. Arvanitopoulos, M. Antoniou, M. R. Jennings, S. Perkins , K. N. Gyftakis, P. Mawby, N. Lophitis; A Defects-Based Model on the Barrier Height Behavior in 3C-SiC-on-Si Schottky Barrier Diodes, IEEE J.Em. Sel Topics Power Electr., 8 (2020) 54